\documentclass[epj,twocolumn]{webofc}
\usepackage[varg]{txfonts}   
\woctitle{The Innermost Regions of Relativistic Jets and Their Magnetic Fields}
\begin{document}
\title{Constraints on Blazar Jet Conditions During Gamma-Ray Flaring from Radiative Transfer Modeling}
%
%

\author{Margo F. Aller\inst{1}\fnsep\thanks{\email{mfa@umich.edu}}
        Philip A. Hughes\inst{1}, Hugh D. Aller\inst{1},  \and
        Talvikki Hovatta\inst{2}
 }

\institute{Department of Astronomy, University of Michigan, 817 Dennison Bldg., Ann Arbor, MI 48109-1042, USA 
\and
Cahill Center for Astronomy \& Astrophysics, California Institute of Technology, 1200 E. California Blvd., Pasadena, CA 91125, USA}

\abstract{As part of a program to investigate jet flow conditions during GeV gamma-ray flares detected by Fermi, we are using UMRAO multi-frequency, centimeter-band total flux density and linear polarization monitoring observations to constrain radiative transfer models incorporating propagating shocks orientated at an arbitrary angle to the flow direction. We describe the characteristics of the model, illustrate how the data are used to constrain the models, and present results for three program sources with diverse characteristics: PKS~0420-01, OJ~287, and 1156+295. The modeling of the observed spectral behavior yields information on the sense, strength and orientation of the shocks producing the radio-band flaring; on the energy distribution of the radiating particles; and on the observer's viewing angle with respect to the jet independent of VLBI data. We present evidence that, while a random component dominates the jet magnetic field, a distinguishing feature of those radio events with an associated $\gamma$-ray flare is the presence of a weak but non-negligible ordered magnetic field component along the jet axis.}
\maketitle
\section{Introduction}
\label{intro}
Motivated by the combination of the dominance of blazars in the GeV $\gamma$-ray sky detected by the Large Angle Telescope aboard {\it Fermi} \cite{nol12}, by the broadband spectral information which has  underlined the importance of this emission in the energy budget of these objects \cite{abd10}, and most particularly by increasing evidence supporting a scenario in which this high energy emission arises within the parsec scale jet of these sources (e.g. \cite{agu11,jor13}) at or near to the radio band emission site, we initiated a program to intensively monitor the linear polarization and total flux density variability at centimeter band in a small sample of $\gamma$-ray flaring sources. The radio-band linear polarization observations probe the magnetic field structure in the emitting region, and they can be used to identify changes associated  with the natural development of instabilities in the relativistic flows which can lead to shocks. Such shocks have widely been accepted as the origin of major flares detected in the optical and radio bands since the 1980s \cite{hug85, mar85}. The program data were obtained with the University of Michigan 26-m equatorially-mounted telescope (hereafter UMRAO) operating at  14.5, 8.0, and 4.8 GHz from the commencement of {\it Fermi} cycle 2 in August 2009 until the closure of UMRAO in June 2012. Many highly variable QSOs and LBLs detected by {\it Fermi} had already been included in the UMRAO monitoring program. However, for several sources of interest, the data sampling was insufficient to adequately track the linear polarization variability. The goals of the new program which focused on linear polarization variability were twofold. First, we wished to determine whether we could find evidence for the presence of shocks in the radio-band data during at least some of the $\gamma$-ray flares detected by {\it Fermi}, and secondly, we expected to use the data in combination with new radiative transfer simulations incorporating propagating shocks to constrain the physical conditions within the radio jet at or near to the presumed emission site of the $\gamma$-ray flares. Many of the jet properties which impact studies of the emission process in the $\gamma$-ray band have been
inadequately constrained for key objects. Relevant properties include the energy distribution of the emitting particles, the Doppler factor of the flow, the degree of order and topology of the magnetic field in the emitting region, and the character of the shocks including their strength and sense (forward or reverse); these are expected to play  a role in the production of at least some of the $\gamma$-ray flares (e.g. \cite{der09}). 
\section{The Data: Source Selection and Observing Procedures}
\label{sec-2}
The data used for our analysis consist of linear polarization and total flux density observations at three centimeter-band frequencies, and it is the temporal coverage allowing us to identify the range of change at the three frequencies which provides the main model constraints. The sample objects were selected because they were expected to be bright and variable in {\it both} the radio band and at GeV energies based on their variability histories. Radio-band total flux densities of a few hundred milli-Janskys are generally required to obtain adequate signal-to-noise in the UMRAO linear polarization data since the fractional linear polarization is typically only a few percent. Our original sample was comprised of about 30 blazars, but this number was subsequently reduced to about 18-20 blazars in order to follow the variations in the most active sources with higher observing cadence. The time scales of individual linear polarization flares are generally on the order of several weeks to a few months. All of the targets are in the MOJAVE 15 GHz imaging program, providing complementary information on changes in the source structure at a frequency which overlaps our single dish monitoring observations; in general the MOJAVE cadence is only several times per year, and the observations are not optimized for following changes in polarization structure. The general UMRAO observing and reduction procedures used for this work are described in detail in \cite{all85}. In brief, each daily-averaged UMRAO observation consists of a series of measurements over a 25 to 45 minute time period. Source observations are interspersed with observations of calibrators every 1 to 3 hours to monitor the antenna gain and to verify the telescope pointing; hence the number of targets which could be monitored in a 24-hour run was limited to only 20-24 sources. 

\begin{figure*}
\centering
\includegraphics[scale=0.50]{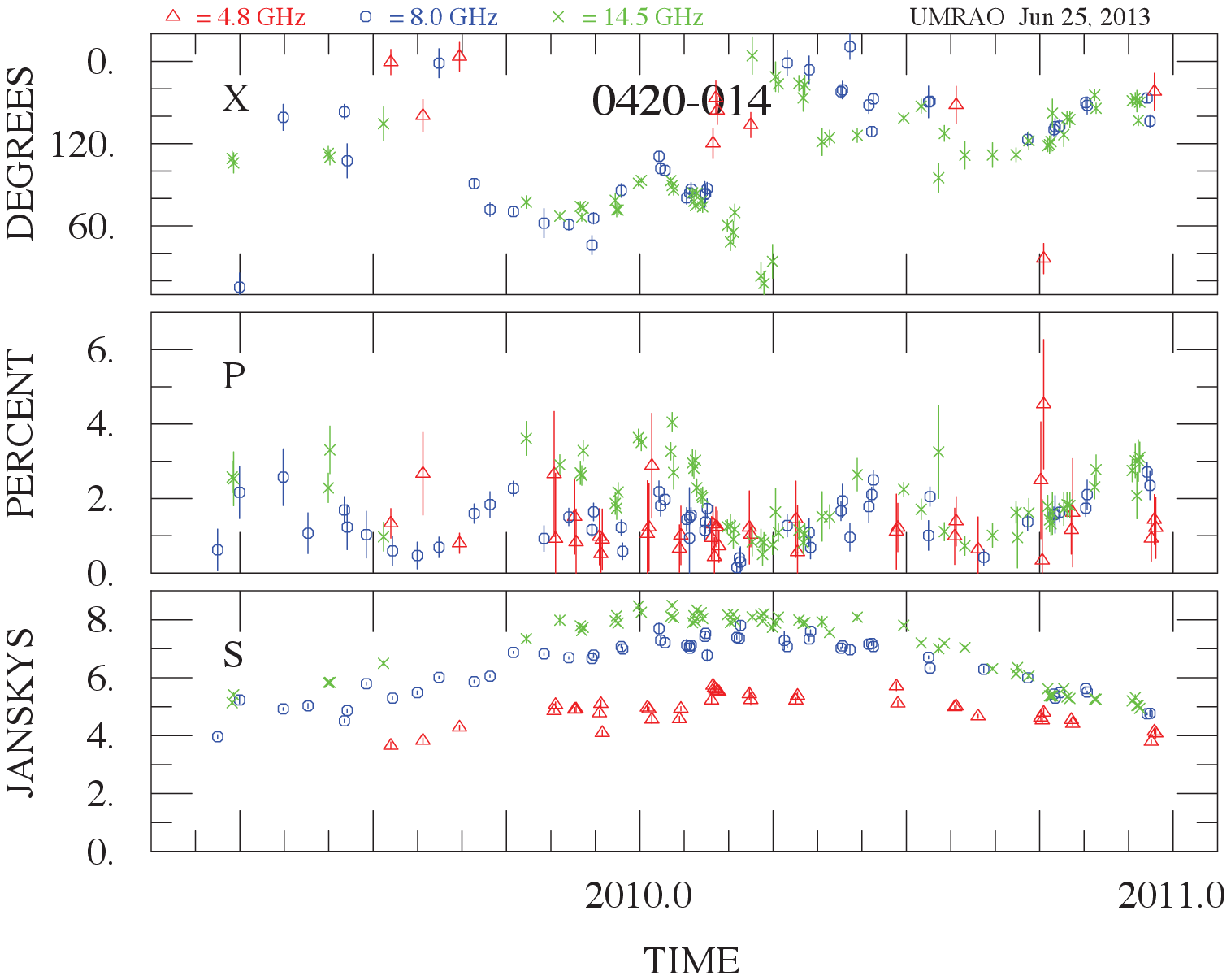}
\includegraphics[scale=0.32]{0420_10step.eps}
\caption{Comparison of the observed and simulated light curves for the 2009 - 2010 event in 0420-014. Left: From bottom to top the daily-averaged UMRAO observations of the total flux density, the fractional linear polarization, and the EVPA are shown. The range of the EVPAs in the UMRAO data is restricted to 180$^{\circ}$ due to ambiguities in the determination procedure. Right: The simulated light curves. These computations have been carried out at 3 harmonically-related frequencies separated by $\sqrt{3}$ which correspond to the UMRAO observing frequencies of 14.5, 8.0, and 4.8 GHz. The symbols for the three frequencies follow the convention adopted for the UMRAO data. The abscissa shows fractional time during the event modeled.\label{fig-1}}
\end{figure*}

\section{ The Model}
\label{sec-3}
\subsection{The Framework for the Simulations}
In \cite{hug11} we described the development of models for simulating the time-dependent radio-band synchrotron emission from blazars.  These radiative transfer models incorporated a propagating shock. We assume the presence of a passive, turbulent magnetic field represented by cells with  randomly-oriented field direction before the passage of the shock. With the passage of the shock the magnetic field is compressed, and the degree of order is increased, and there is an increase in the particle density and hence the emissivity. While this magnetic field is assumed to be predominantly disordered in our scenario, a fraction of the magnetic energy is assumed to be in an ordered component in order to produce a well-defined electric vector position angle (hereafter EVPA) in the quiescent state. This assumption is supported by the low levels of fractional linear polarization identified during relatively quiescent states in the UMRAO data for a few sources exhibiting a well-defined base level during relatively inactive time periods. The propagating shock (or shocks) is allowed to be oriented at any direction to the flow direction, and it is not restricted to a transverse orientation relative to the flow as in our earlier modeling \cite{hug85}. To simplify the computations, each shock is assumed to span the entire cross section of the flow, and multiple shocks contributing to a single outburst are assumed to have the same orientation. This orientation, relative to the jet outflow, is specified by two angles; these are the shock obliquity, $\eta$, measured with respect to the upstream flow direction and the azimuthal direction of the shock normal, $\psi$. However, early explorations of parameter space showed that the simulations are relatively insensitive to changes in the azimuthal direction \cite{hug11}. The computations assume that each shock propagates at a constant speed and that subsequent shocks contributing to an event do not shock pre-shocked plasma. The shocked flow itself is specified by a length (l) defined in \cite{hug11}, a compression factor ($\kappa$), and a start time (t$_o$).
\begin{figure*}
\centering
\includegraphics[scale=0.52]{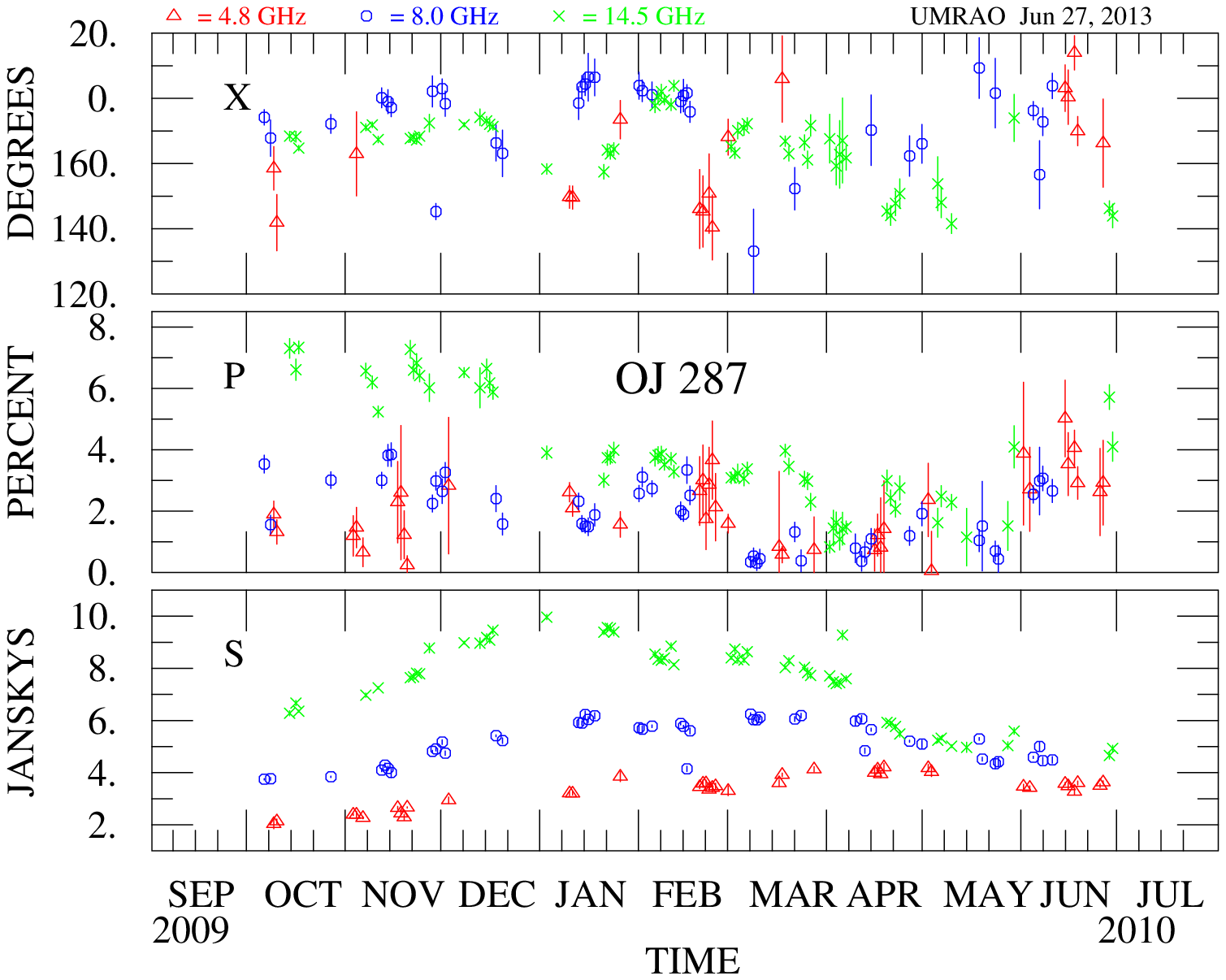}
\includegraphics[scale=0.35]{OJ287_10step.eps}
\caption{Comparison of the UMRAO (left) and the simulated (right) light curves for OJ~287 during the ten-month event modeled. The $\gamma$-ray flare occurred in fall 2009. The dominant characteristic of  the total flux density light curve is the inverted spectrum during the rise portion of the outburst, followed by a spectral flattening during the decline. The spectral evolution in both the linear polarization and in the total flux density is very well-reproduced by this simulation. The figure conventions are as in Figure~\ref{fig-1}.\label{fig-2}}
\end{figure*}

\subsection{Constraints on the Free Parameters}
While the simulations require input values for several `free' parameters specifying the character of the flow and the shock parameters, these are, in fact, very well-constrained by using the character of the spectral evolution of {\it both} the linear polarization and the total flux density.  Initial estimates for the observer's viewing angle were adopted based on results taken from the literature from past VLBI studies. The relevant free parameters and the most important UMRAO constraining data for each of them are summarized in Table~\ref{tbl-1}. An additional parameter, the fraction of the energy density which is in an ordered component of the magnetic field, is discussed in Section~\ref{sec-4}.
\begin{table}
\centering
\caption{Free Parameters and Data Constraint}
\label{tbl-1}       
\begin{tabular}{ll}
\hline
Parameter            & UMRAO constraint \\
\hline
Cutoff LF (energy)    &  spectral behavior    \\
Bulk LF (flow)       &  P\%                    \\
Shock Sense (F or R) &  light curves        \\
Shock length (l)     &  flare shape  (S)       \\
Azimuthal angle ($\psi$)  &  UMRAO P\%        \\
Shock Compression ($\kappa$)  & $\Delta$S and P\% \\
Shock Obliquity ($\eta$)   &  $\Delta$EVPA        \\
Observer's Viewing Angle       & Peak P\%         \\         
\hline
\end{tabular}
\newline Note to table: LF denotes Lorentz factor
\end{table}
\section{Results from Comparison of the Data and the Simulations}
\label{sec-4}
The modeling has now been carried out for single radio-band events with large, temporally-associated $\gamma$-ray flares in three sources. While the time scales for the variability in the radio and $\gamma$-ray flares are characteristically different (typically hours to weeks at high energies and from months to years in the radio band; see \cite{all10}) often hampering unambiguous associations of specific flares across bands even when delays due to opacity effects are considered, we have based our cross-band associations on the presence of a rise in the $\gamma$-ray photon flux by  a factor of several above the baseline level during the rise phase of a radio band total and polarized flux outburst. An association between the radio-band flare phase and the occurrence of a physically-related $\gamma$-ray flare has been proposed since the first EGRET detections of $\gamma$-ray flaring in the 1990s based on statistical studies \cite{lah03}. Further we selectively chose light curves dominated by only a few, strong events. All three target sources are included in {\it both} the MOJAVE (15 GHz) and the Boston University (43 GHz) imaging programs. These programs have provided complementary information on changes in the source structure associated with the $\gamma$-ray flares, the range of pattern speeds in the jet of each source, the projected jet orientation at individual epochs, and temporal changes in this orientation relative to the direction of the observer; these are useful for interpretation of our own results. The trio of sources represents a range of source properties which potentially impact the emission properties seen by the observer: these include the  optical class (2 QSOs; one BL Lac object), the source redshift (0.306$\leq$z$\leq$0.9161), and the  maximum apparent component speed in the parsec scale region of the jet based on the MOJAVE data (5.74c$\leq\beta_{app}\leq$24.59c \cite{lis13}). 

A complication in carrying out our analysis is that in general the centimeter-band light curves are blended due to the overlapping of individual flares in the single dish monitoring data. In order  to resolve the individual flares, we developed a method which uses a combination of  the structure apparent in the total and polarized flux density light curves and the theoretically-expected burst profile shape based on our simulations for a single shock.  Our procedure differs from previous work by other investigators
which often assumes a generic exponential shape (e.g. \cite{leo11}).
The goal of our work is to determine whether we can reproduce the general characteristics of the observed light curves with our adopted scenario incorporating these individual shocks. While we have explored a range of parameters in our modeling, at this stage of the work we have not attempted to formally maximize the fits of the simulations described below to the observed light curves during the events modeled. We caution that our model adopts a relatively simple approach, while real flows are complex hydrodynamically. Jet curvature on a variety of scales in particular can influence the observer's perception of the observed variability.
\subsection{0420-014 (PKS 0420-01)}
In Figure~\ref{fig-1} we compare the data to the simulation based on the flow and shock parameters listed in Table~\ref{tbl-2} during $\gamma$-ray flaring detected by the {\it Fermi} LAT which peaked in late January 2010. Note that the simulations successfully reproduce the general characteristics of the observed light curves. These features include the range of change in the total flux density at 14.5 GHz
(4-8 Jy), the spectral evolution of the total flux density (from flat to inverted to flat) during the event modeled, and the range of change of the fractional linear polarization (0-3\%). While the behavior of the EVPAs is complex, the range of the change, $\Delta_{EVPA}$, is consistent with the passage of transverse shocks. Three shocks were required to reproduce the overall character of the observed event.
\begin{figure*}
\centering
\includegraphics[scale=0.53]{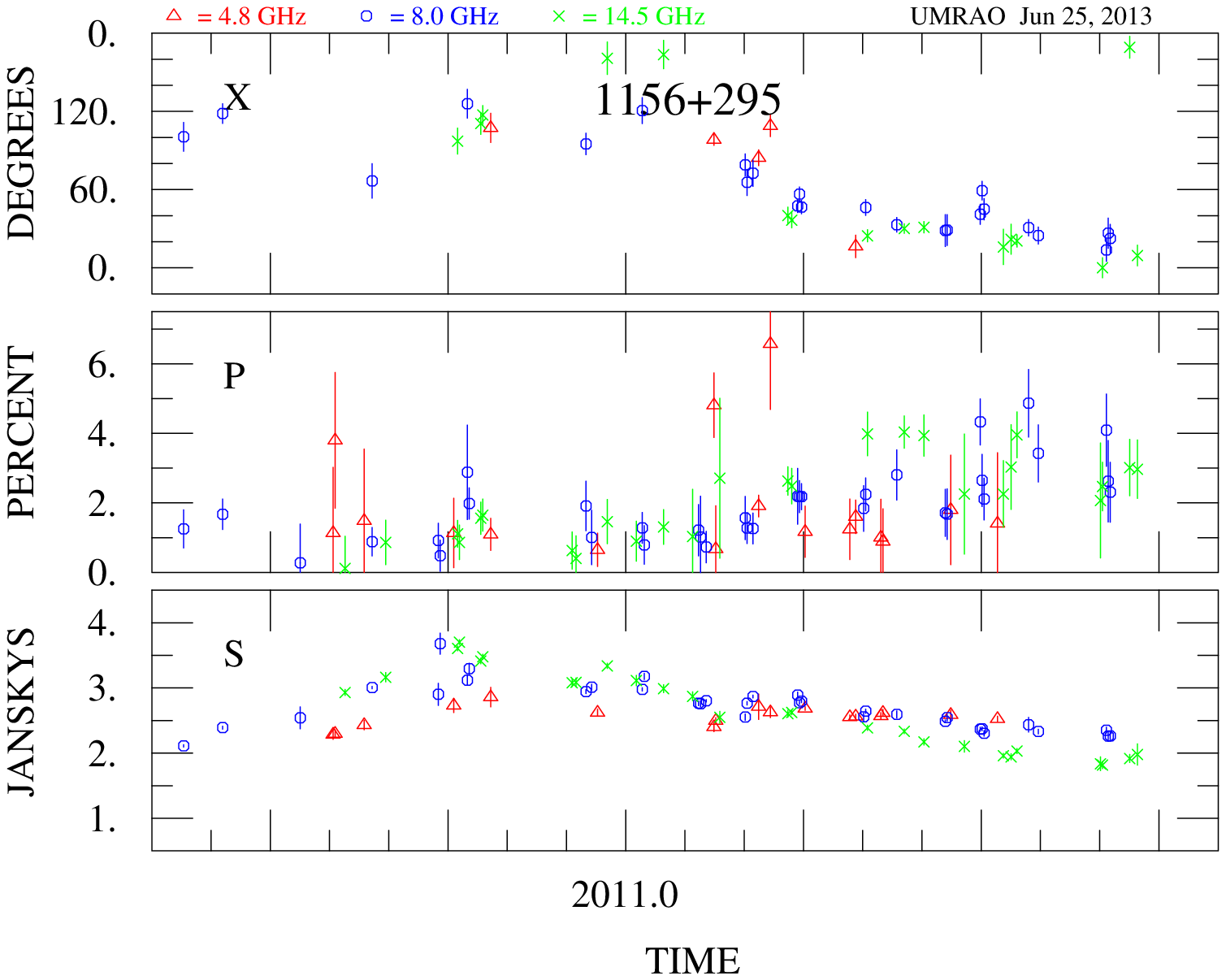}
\includegraphics[scale=0.33]{1156_10step.eps}
\caption{Comparison of the UMRAO  and the simulated light curves for 1156+295 during a series of $\gamma$-ray flares which occurred in 2010, peaking in August. 
The time window of the data shown is 2010.35 -- 2011.75. The figure conventions are as in Figure~\ref{fig-1}. \label{fig-3}}
\end{figure*}

The VLBI structure at 43 GHz has been analyzed in detail during the time window of our analysis, and results were presented in a paper at this meeting \cite{tro13}. While these data probe a region of the jet upstream of the 15 GHz emission site, it is instructive to compare the ejection times of these components with our adopted shock start times for the simulation shown; this comparison is discussed more fully in \cite{all13a}. We find that the ejection times for components K3 and K4 are in excellent agreement with the start times of our second and third shocks. For K3 the knot ejection time is 2009.444$\pm$0.110, and the corresponding shock start time is 2009.6; for K4 the knot ejection time is 2009.892$\pm$0.030, and the shock start time is 2009.95. A component ejection time has been provided privately by the authors for component K2. While the error bar associated with this ejection time is large, the value within the errors is an acceptable match to the start time of our first shock. Hence there is a temporal association between all three model shocks contributing to the emission from this event and the ejection of new VLBI components from the millimeter-band core in the resolved inner jet observations. 
\subsection{OJ ~287 (0851+202)}
In Figure~\ref{fig-2} we compare the data and the simulations for OJ~287 during a large radio-band outburst which commenced in September 2009 just after the source emerged from its annual sun gap. This outburst lasted until mid-2010. The corresponding $\gamma$-ray flare began in September 2009 and peaked in late October 2009. The emission during this event is self-absorbed as shown by the inverted total flux density spectrum. The structure apparent in the light curves is consistent with the passage of three propagating oblique shocks with a shock obliquity of 30$^{\circ}$. A simulation produced with the parameters shown in the table is able to reproduce the range of values  in the total flux density light curve in the bottom panel (2-9 Jy) and the spectral character during the evolution of the event (self-absorbed during the outburst rise and maximum; flattening during the outburst decline) as well as the range of the variation in the fractional linear polarization (0-6\%). The range of the swing in the  EVPAs is also reproduced by the simulation shown. However, the detailed spectral behavior during this event is complex, and the details of the simulation show some differences when compared with the data.

 As in the case of 0420-014, it is useful to compare the modeling results with independent VLBI results. The Lorentz factor of the modeled shocks is about 20, with  little dispersion in this parameter since the value for the compression is similar for the three shocks. Note that this indicates the speed of the shock transition and not the flow speed. From the modeling the value of $\beta_{app}$ is found to range from about 5c to about 20c as the viewing angle is changed from 0.5 to 2.5$^{\circ}$. This range in values is consistent with the range in observed apparent component speed identified both from the MOJAVE 15 GHz data \cite{lis13} and from the 43 GHz data from the BU program \cite{agu12} for this well-studied object. Because of the importance of viewing angle on the derived value of $\beta_{app}$, we suggest that this range in the VLBI values of the apparent component speed may arise because different flow segments are observed at slightly different angles during the sequence of VLBI observations. This may occur  either because the data at each epoch probe a different part of the flow or because the ejection angle itself changes slightly from epoch to epoch.
\subsection{1156+295 (TON 599)}
In Figure~\ref{fig-3} we compare the observed and simulated light curves for the QSO 1156+295. A series of intense $\gamma$-ray flares commenced in  April 2010 which peaked prior to the peak in the radio-band event shown in the figure.  The simulation (right plot in figure)  incorporates four transverse propagating shocks and assumes that 50\% of the energy density in the magnetic field  is in an ordered component oriented along the jet axis. The latter is discussed more fully below. The simulation reproduces the peak value, the spectral rise, and the relative duration of the decline portion of the outburst in total flux density; the monotonic rise in fractional linear polarization during the decline portion of the total flux density profile and the approximate  peak fractional linear polarization; and the EVPA swing through about 90$^{\circ}$ characteristic of a transverse shock and the general spectral behavior of the EVPAs during the time period modeled. While a range of values of the apparent speed of the components, $\beta_{app}$, has been identified by a recent MOJAVE study \cite{lis13}, the fastest of these speeds at 24.6c is in good agreement with the value of 22c found from the modeling.
\section{Source Properties Extracted Using the Radiative Transfer Simulations}
\begin{figure}
\centering
\includegraphics[width=6.5cm,clip]{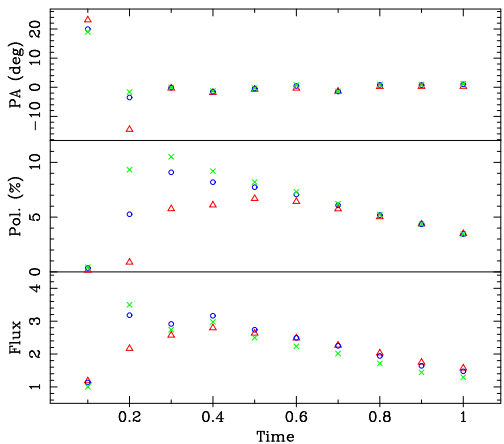}
\caption{Simulated light curves for 1156+295 assuming the weakest axial B field able to give a well-defined EVPA in the quiescent state. Note that in this simulation the peak amplitude of the fractional linear polarization is near 10\% at 14.5 GHz and that the EVPAs are flat at nearly the same value for all three frequencies. This behavior does not match the spectral evolution observed for this source as shown in Figure~\ref{fig-3} (left).}
\label{fig-4}
\end{figure}

In Table~\ref{tbl-2} we summarize the results of our analysis for the events modeled. An unexpected outcome of the work is the realization  that a substantial fraction  of the magnetic field energy density must be in an ordered component along the jet axis.  This result first became apparent during the analysis of the final source in our trio, 1156+295. Without inclusion of this ordered component, as shown in Figure~\ref{fig-4}, the simulated fractional linear polarization was found to exceed 10\%, a result in conflict with the data, and the time variation of the EVPAs was too flat compared to the  observed variability behavior. Initial modeling had already been carried out for PKS 0420-01 and OJ 287 prior to this study (e.g. \cite{all13b}); however, the new simulations shown here, which included the percentages of the energy density in the axial magnetic field listed in Table~\ref{tbl-2}, substantially improved the `fit' of the simulated light curve to the observations. 

\section{Discussion}
Both the validity of our propagating shock scenario and the viability of our method for obtaining source properties based on a comparison of linear polarization variability data and radiative transfer simulations have been demonstrated in this pilot study. We plan to analyze additional events in future work. This new work will include the analysis of additional events in the same three sources in order to look for changes in characteristics from event to event, and the modeling of events in other program sources to explore the range of values of the properties listed in Table~\ref{tbl-2} for  $\gamma$-ray flaring sources.  In the three blazars studied to date the shocks were found to be forward moving; this sense is in contrast to our earlier work in the 1980s where the shocks were found to be reverse. However, reverse shocks led to values of the Doppler boost and apparent motion which are in disagreement with the excellent VLBI data obtained during the events modeled, and this shock sense was rejected on those grounds.  The simulated linear polarization light curves are very sensitive to the choice of observer's viewing angle for fast flows seen at orientations of only a few degrees as in these three blazars with nearly line-of-sight viewing angles. Consequently our linear polarization data are a powerful constraint on this important parameter and a means, independent of VLBI observations, to determine it.  We note that viewing angle determinations obtained from VLBI studies are sensitive to the segment of the flow dominating the emission at the observing frequency used for the measurement and hence suffer from this inherent bias. An additional parameter identified from our modeling of these events is the low energy cutoff of the energy distribution of the radiating particles. In the case of OJ~287, this was found to be lower than the values identified in the other two sources analyzed here. An important source property identified through the modeling is the presence of a substantial ordered axial magnetic field in these $\gamma$-ray flaring sources. 

Overall, our method has yielded results consistent with independent VLBI determinations while providing more detailed information about the flow conditions where the flaring is produced. Although in this exploratory phase of the work we have restricted our analysis to sources which are well-observed in VLBA programs for comparison purposes and to narrow parameter space for some properties, in principal the method can be applied to sources which are not observed intensively with the VLBA but for which single-dish, multifrequency, radio band polarimetry data are available or will be available. This use emphasizes the importance of obtaining such polarimetry data in future programs.

\begin{table}
\centering
\caption{Summary of Derived Source Properties}
\label{tbl-2}       
\begin{tabular}{llll}
\hline
Parameter            & 0420-014 & OJ 287 & 1156+295 \\
\hline
Cutoff LF (energy)   & 50      & 10      &   50    \\
Bulk LF    & 5.0     & 5.0    &    10.0    \\
Shock Sense          &  F      & F      &     F      \\
Number of Shocks     &  3      & 3      &   4        \\
Shock Obliquity      & 90$^{\circ}$   & 30$^{\circ}$   & 90$^{\circ}$  \\
Viewing Angle        & 4.0$^{\circ}$    & 2.5$^{\circ}$  & 2.0$^{\circ}$  \\
Axial B Field        & 16\%           &   50\%         & 50\%           \\
\hline
\end{tabular}
\end{table}

\vskip 0.5 cm
\section{Acknowledgements}
This work was supported in part by NASA Fermi GI grants NNX09AU16G, NNX10AP16G, NNX11AO13G. and NNX13AP18G. It has made use of VLBI data from the  MOJAVE database that is maintained by the MOJAVE team \cite{lis09}. T.H. gratefully acknowledges support from the Jenny and Antti Wihuri foundation. 

%
%

\end{document}